# Extensions around $z = 2$ QSOs

Itziar Aretxaga, B.J. Boyle, Roberto J. Terlevich
*Royal Greenwich Observatory, Madingley Road, Cambridge CB3 OEZ, U.K.*

**ABSTRACT**
Based on an R-band imaging survey of 6 high redshift ($z \sim 2$) and high luminosity ($M_B \lesssim -28$ mag) QSOs, we report the detection of extensions in two radio-quiet and one radio-loud QSO. The extensions are most likely due to the host galaxies of these QSOs, with luminosities of at least $3-7$ per cent of the QSO luminosity. The most likely values for the luminosity of the host galaxies lie in the range $6-18$ per cent of the QSO luminosity.

## 1. INTRODUCTION

The study of high redshift ($z \sim 2$) QSOs offers a unique opportunity to investigate conditions in the early universe. In the currently favoured cold dark matter cosmogony, this epoch corresponds to the period when normal galaxies formed through hierarchical coalescence (Carlberg 1990), thereby giving rise to enormous concentrations of gas in the center of the galaxies, which could feed a central black hole (Rees 1988) or provoke a massive starburst episode (Terlevich & Boyle 1993). As such, this picture is consistent with the observation that QSOs exhibit a peak in their activity at these redshifts (Schmidt, Schneider & Gunn 1991). Indeed, there has been much observational effort to obtain deep multicolor images of high redshift radio galaxies and radio-loud QSOs with spectacular results: enormous structures of up to 100 kpc aligned with the radio-jet (Chambers et al. 1987), luminosities and sizes of the parent galaxies larger than those of normal elliptical galaxies (Lehnert et al. 1992). However, radio-loud quasars are only a small fraction ($< 5\%$) of all QSOs, and many of the conclusions drawn from the studies of radio-loud objects might be unrepresentative of the conditions in the early universe. It is, therefore, important to look into the properties of radio-quiet QSOs, as a better indicator of the general properties of galaxies at high redshifts. We present here deep images of a sample of 6 high redshift, predominantly radio-quiet QSOs. Section 2 gives a summary of the observations, section 3 explains the technique used to detect extensions around the QSOs, and section 4 discusses the likelihood that the extensions are host galaxies.

## 2. OBSERVATIONS

Four radio-quiet and two radio-loud QSOs were observed during 1994 August 8/9 and 9/10 in the Harris $R$ passband with the f/11 auxiliary port of the 4.2m William Herschel Telescope (WHT) of the *Observatorio de Roque de los Muchachos* in La Palma. Luminous ($M_B \leq -28$ mag)* high redshift ($1.8 \leq z \leq 2.2$) QSOs which lay close ($20 \lesssim \theta \lesssim 60$ arcsec) to stars of similar brightness, were selected from the Véron-Cetty and Véron Catalogue (1994) catalogue. The existence of a nearby star allowed us to determine an accurate point spread function (PSF) for each CCD frame. The observations were performed under excellent seeing ($< 0''.9$) in dusty and smoky non-photometric conditions.

We carried out the observations with a $1024^2$ Tektronix CCD chip (TEK1), giving a spatial resolution of 0.105 arcsec/pixel over an unvignetted 1.8 arcmin diameter field. The configuration was chosen to provide an optimal sampling of the PSF, even in the best seeing conditions ($\approx 0''.5$). The chip was readout in QUICK mode, which yields a read noise of 4.3 e$^-$ and a gain of 1.8 e$^-$/ADU.

Each QSO field was observed in a series of 600–900s exposures to avoid saturation of the QSOs and PSF stars. Individual exposures were offset from one another by $\sim 10$ arcsec to prevent the repetition of any coherent signal (sources, cosmic rays, bad pixels) at a particular location on the chip. Table 1 summarizes the observations.

## 3. DATA REDUCTION AND ANALYSIS

The data were reduced using the IRAF software package. For each separate QSO field, the images were first bias subtracted and then flat-field corrected using the sky flat-field derived from the median of all individual frames, once bright stars, galaxies and QSOs (and their surrounding regions) had been masked from the original data.

---

* $H_0 = 50$ Km s$^{-1}$ Mpc$^{-1}$  and $q_0 = 0.5$ are used throughout this work.

**Table 1.** Summary of observations

| Name | B[*] | $M_B$[*] | $z$[*] | Radio-loud?[*] | Exposure time (seconds) | No. of frames | FWHM[†] (arcsec) |
|---|---|---|---|---|---|---|---|
| UM 281 | 17.4 | −28.9 | 1.878 | N | 6600 | 11 | 0.7 |
| 1630.5+3749 | 18.3 | −28.3 | 2.037 | N | 10500 | 15 | 0.6 |
| PKS 2044−16 | 17.4 | −29.1 | 1.943 | Y | 5400 | 9 | 0.9 |
| PKS 2134+008 | 16.8 | −29.6 | 1.936 | Y | 6300 | 7 | 0.6 |
| Q 2230+02324 | 18.0 | −28.7 | 2.150 | N | 5400 | 9 | 0.6 |
| Q 2244−0105 | 18.0 | −28.6 | 2.040 | N | 5400 | 9 | 0.7 |

[*] B, $M_B$, $z$ and radio properties from Véron–Cetty and Véron (1994).
[†] Stellar FWHM measured on co-added frame.

We then applied the PSF subtraction techniques provided by the IRAF package DAOPHOT to establish whether QSOs in this study exhibited any significant non-stellar extensions to their profiles. This procedure has already been successfully used to isolate intervening galaxies in the line of sight of QSOs (Steidel & Hamilton 1992). For each individual frame, we defined a 2-dimensional PSF using the brightest of the closest stellar companion to the QSO. The PSF was defined over 30–40 pixels in radius. We then scaled the PSF to match the luminosity of the QSO and other nearby stars over the same region, and subtracted the scaled PSF from them. The remaining residuals in the non-PSF stars provided an accurate check of the validity of the subtraction process. We accepted the PSF subtraction if the residuals in the non-PSF stars accounted for less than $1\sigma$ of the Poisson noise expected from the subtraction technique, and showed no systematically repeatable pattern from frame-to-frame. The $1\sigma$ level was calculated by combining the Poisson noise from the different sources (sky subtraction, PSF fitting, QSO profile) in the PSF subtraction process. For one field, Q2230+0232, there was just one bright star near the QSO which was used to model the PSF. Although the residuals for the non-PSF stars are consistent with zero, their distance from the field centre ($> 40$ arcsec) results in a distinctive pattern in their residuals which repeats from frame to frame. For PKS 2044−168 there were two bright stars near the QSO, but neither star yielded a PSF which produced a non-zero residual when subtracted from the other and, therefore, we cannot assess the significance of the residuals left in the QSO position.

For the remaining fields the subtraction of non-PSF stars was successful. The subtracted frames were registered using the relative frame-to-frame X,Y shift derived from the positions of three bright stars in each original frame, and co-added to produce a final subtracted image. The same transformations were used to register and co-add the original frames to produce a final un-subtracted image.

The left panels in figure 1 show the fields of the QSOs UM 281, 1630.5+3749, PKS 2134+008 and Q2244−0105. The right panels in figure 1 show the same fields after subtracting the corresponding PSFs from both QSOs and stars. Figure 2 shows in bigger detail $6'' \times 6''$ sections of the residuals left in the positions of the QSOs and non-PSF stars. The residuals left in the position of the stars are consistent with zero at a $1\sigma$ level (see above). Table 2 summarizes the significance of the residuals in the number of ADU. We have detected residuals in excess of $3\sigma$ in the following QSOs: 1630.5+3749 ($4\sigma$), PKS 2134+008 ($3\sigma$) and Q 2244−0105 ($3.7\sigma$). All the residuals show a 'doughnut' shape with a well of negative counts in the centre. This indicates that there is a flatter component below the PSF in the centres of the QSOs, from which the nuclear (PSF) contribution has been over-subtracted. To estimate the true luminosity of these systems, we have subtracted smaller amounts of the PSF in order either a) to produce zero counts in the center of the residuals or b) to achieve a flat-top profile with no depression in the center. We regard these quantities as a lower limit and a best estimate, respectively, of the total luminosity of these extended components. In the lower limit case, the luminosities of the extensions comprise 3–7 per cent of the luminosity of the QSOs. In all cases, the FWHM of the flat-top residual profiles are significantly larger than the FWHM of the stars in each field (see tables 1 and 3). The estimates of the extended component luminosities derived from flat-top profiles increase to 6–18 per cent of the luminosity of the QSOs. To obtain a crude zero-point ($\pm 0.3$ mag) for the images, we used the E magnitudes of the bright stars in the field provided in the APM Northern Sky Catalogue (Irwin, Maddox & McMahon 1994), which correspond to R band (Evans 1989). Since we do not know the spectral energy distribution of the extensions, we have chosen not to correct for spectral slope or bandpass in these magnitudes, so that they correspond to magnitudes in a rest-frame band centred at $\sim \lambda 2300$ Å. We list these magnitudes in table 3.

To check the reliability of the luminosities obtained from the PSF subtraction technique, we tried to recover the luminosities of known faint compact (FWHM$\sim 1 - 2$ arcsecond) galaxies in each CCD frame after we had 'added in' scaled versions of the stellar PSF, mimicking the contribution from the unresolved bright QSO nucleus. We found that subtractions which produced flat-topped residuals could recover the input luminosity of the galaxies within an error of less than 4 per cent, if the galaxies contributed more than 3 per cent to the total luminosity of the galaxy+PSF system. For galaxy contributions of $\lesssim 1$ per cent, the input luminosity of the galaxies were poorly recovered, with errors of up to 30 per cent.

## 4. DISCUSSION

Out of a sample of one radio-loud and three radio-quiet QSOs with suitable PSF stars, we have been able to detect extensions in three cases. The best estimates for luminosities of these systems lie between 6–18 per cent of the total QSO luminosity. Host galaxies of radio-loud QSOs at $z \sim 2$ which comprise up to 20 per cent of the QSO R band luminosity have already been detected by Lehnert et al. (1992) but, to date, no other detection has been made of hosts of high-redshift radio-quiet QSOs.

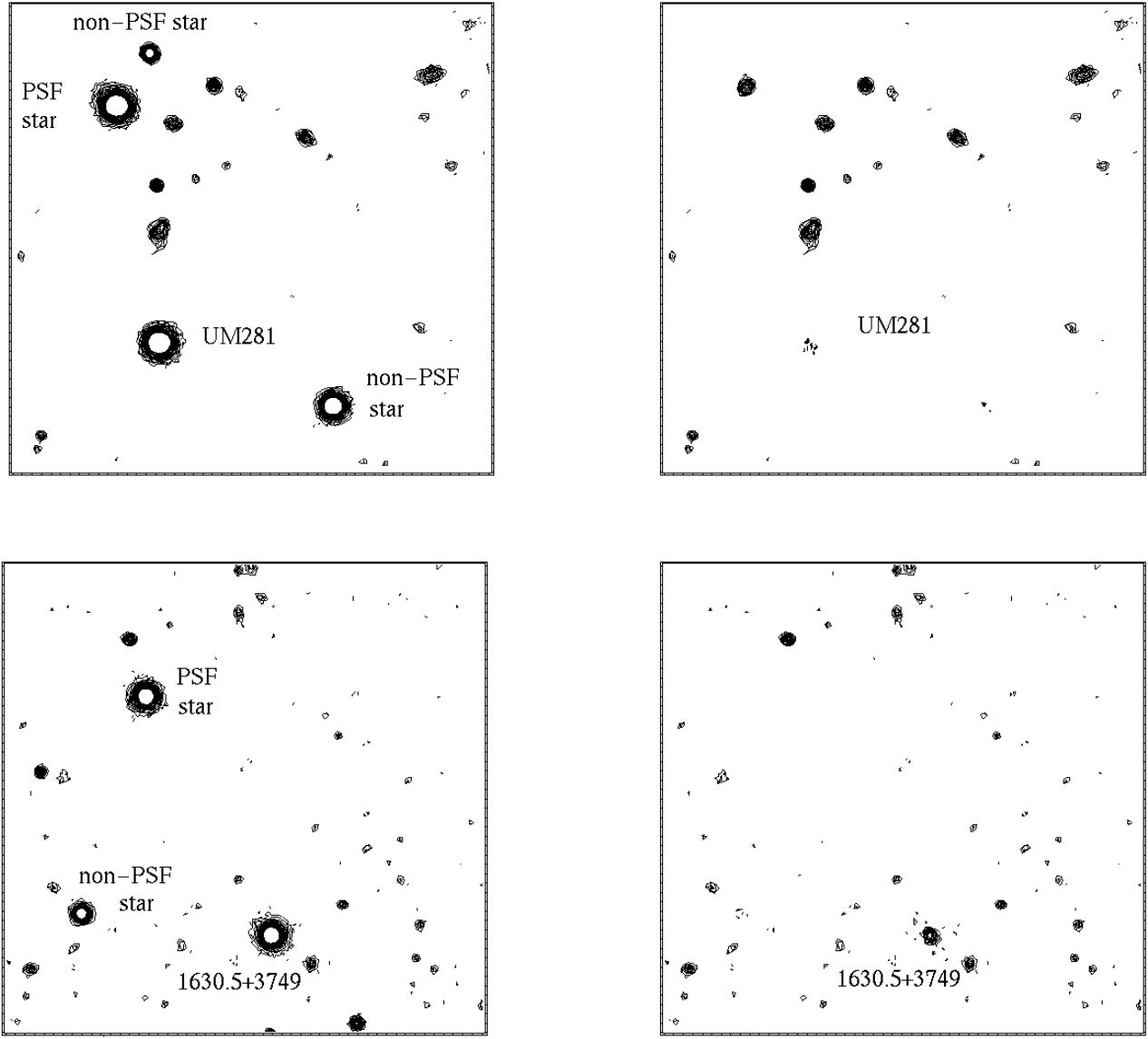

**Figure 1.** $50'' \times 50''$ fields of the QSOs UM 281, 1630.5+3749, PKS 2134+008 and Q 2244−0105. Left panels show the fields before subtracting the PSF stars. Right panels show the residuals left in the position of the QSOs and stars, after a luminosity scaled PSF has been subtracted. QSO residuals show a 'doughnut' profile, with negative counts in the center. The minimum contour level has been set to be $\sim 2\sigma$ above the sky level. North is left, and East is down.

**Table 2.** Significance of the 'doughnut' residuals: number of ADU

| Field | QSO | PSF star | non-PSF stars |
|---|---|---|---|
| UM 281 | 2123 ± 887 | 830 ± 890 | 740 ± 754 |
| 1630.5+3749 | 3932 ± 1003 | 491 ± 999 | 522 ± 733 |
| PKS 2134+008 | 7940 ± 2539 | −1491 ± 2207 | 1706 ± 1791 |
| Q 2244−0105 | 4997 ± 1370 | 164 ± 1264 | −324 ± 1071 |

For radio-loud QSOs, the red colours of the extensions (Lehnert et al. 1992) favour a host galaxy origin of the light, rather than scattering by dust or electrons in the halo of the QSOs. Although we lack colour information, the similarility between the nuclear to extended component luminosity ratio in radio-loud QSOs with that observed here, leads us

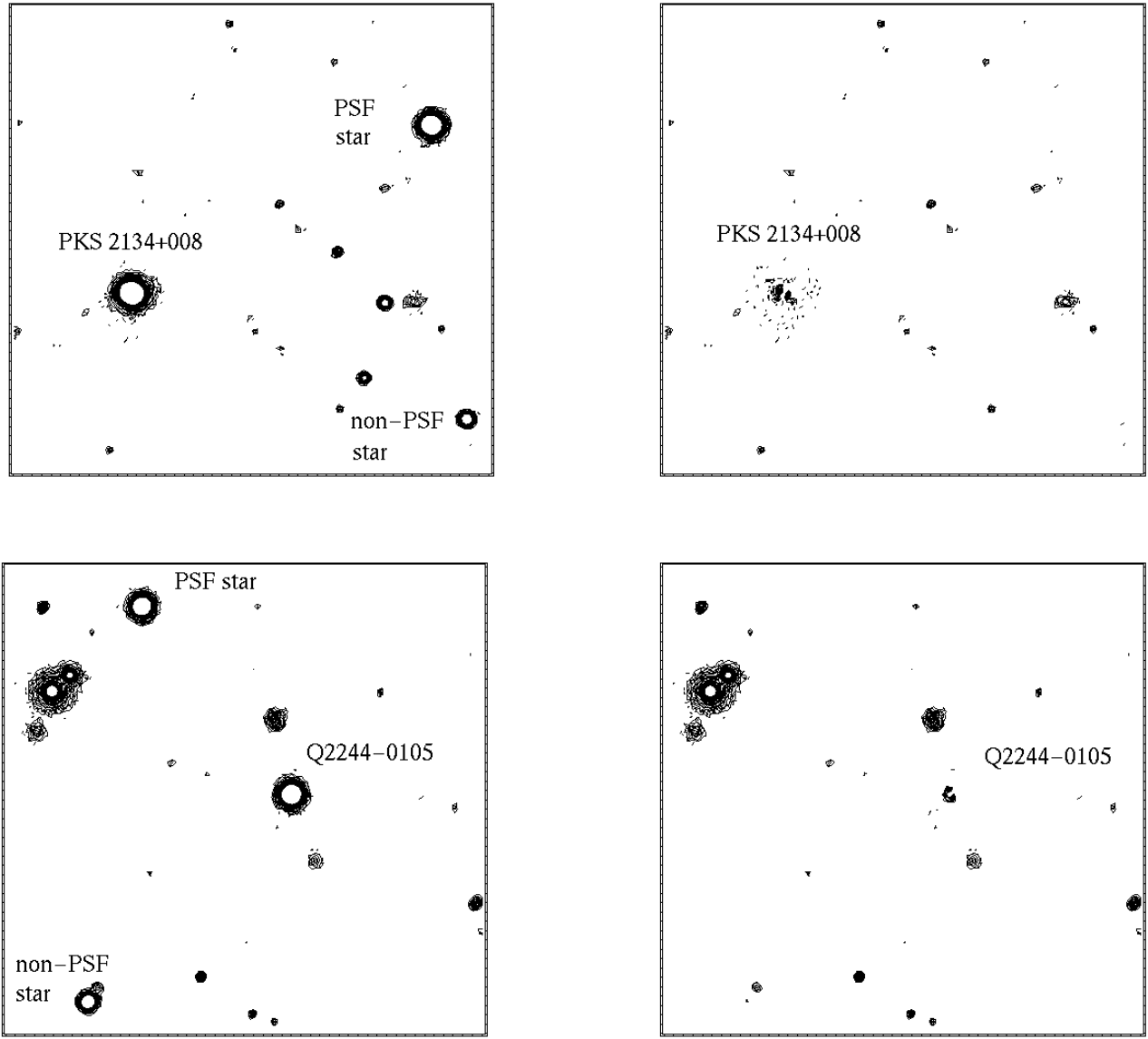

**Figure1:** (continued)

Table 3. Magnitudes of the QSOs and their extensions

| Name | $M_R^*$ (QSO) | $M_R^1$ | $R^1$ | $ADU^1$ | $M_R^2$ | $R^2$ | $ADU^2$ | $FWHM^2$ (arcsec) |
|---|---|---|---|---|---|---|---|---|
| 1630.5+3749 | −28.7 | −24.8 | 21.7 | 15170 ± 1001 | −25.9 | 20.9 | 32279 ± 998 | 1.05 |
| PKS 2134+008 | −30.1 | −25.9 | 20.4 | 45509 ± 2536 | −26.6 | 19.8 | 80597 ± 2534 | 0.80 |
| Q 2244−0105 | −29.0 | −25.6 | 20.9 | 22595 ± 1368 | −26.7 | 19.9 | 62975 ± 1365 | 0.84 |

* Total QSO $M_R$ absolute magnitude, including extension.
1 Properties for extension with zero counts in the center after the PSF subtraction.
2 Properties for extension with flat-topped profile after the PSF subtraction.

to believe that the extensions we are detecting are also due to light from the host galaxy. Indeed, scattering models require the presence of a powerful transverse radio-jet (Fabian 1989) which is unlikely to be present in either the radio-quiet QSOs, or the core-dominated (Murphy, Browne & Perley 1993) radio-loud QSO around which we have detected extensions.

An alternative origin for the extensions could be extended nebular continuum, seen to be a major contribution in three powerful radio galaxies (Dickson *et al.* 1995). However, for nebular luminosities of $M_R \sim -26.5$ mag, as im-plied by these observations, the predicted narrow H$\beta$ luminosity would be about $3 \times 10^{44}$ erg s$^{-1}$, nearly half of the luminosity derived to be emitted by the broad component from the PSF light. Under these circumstances, the QSOs would exhibit prominent narrow lines with central peak intensities of more than 3 times those of the broad lines. The spectra of these QSOs do not show such prominent narrow lines (Foltz *et al.* 1989, Osmer *et al.* 1994).

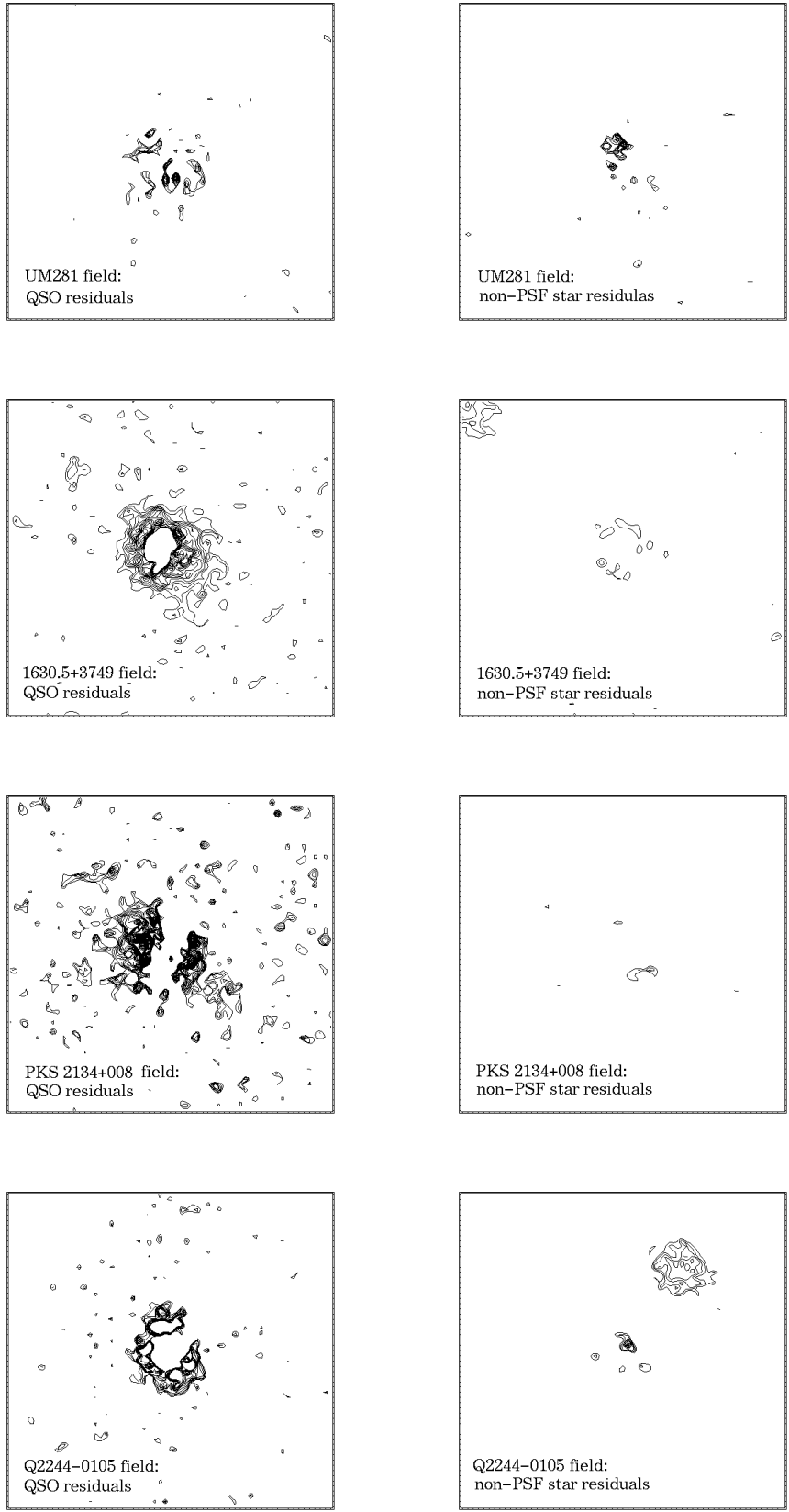

**Figure 2.** $6'' \times 6''$ fields of the residuals of QSOs and non-PSF stars shown in figure 1. North is to the left and East is down. The centroids of the corresponding non-subtracted images lie in the center of the plots. Note that the non-PSF star in the field of Q 2244−0105 has a faint galaxy 1.7 arcsec to the southwest.

Galaxies as luminous as the extensions detected here have already been found in the imaging survey of radio-loud QSOs carried out by Lehnert *et al.* (1992). Four of the objects of their sample, with similar redshift to those in our sample, show 'fuzz' around the PSF of the nucleus. In the observed $R$ frame the absolute magnitude of this 'fuzz' ranges from $-24.8$ to $-26.3$ mag, as derived from the B and K colours they report. Their galaxies are thus $\sim 12$ times more luminous than present-day giant ellipticals and $\sim 2.5 - 3$ mag brighter than the hosts of low redshift QSOs. Our observations show that, if the extensions we have detected are indeed galaxies, extraordinary massive and luminous galaxies are not only a characteristic of radio-loud objects, but of QSOs as an entire class.

Indeed, the radio-loud QSO studied in this sample does not exhibit a significantly larger or a more luminous extension than those of radio-quiet QSOs. One of the radio-quiet QSOs exhibits no significant evidence for any extension. This is in contrast to the work of Lowenthal *et al.* (1995) who have established upper limits to the observed K luminosity of the host galaxies of a sample of high redshift ($z \approx 2.5$) radio-quiet QSOs to be $\sim 2$ mag less luminous than those of radio-loud QSOs (Lehnert *et al.* 1992). A similar result is found in optical studies of low-redshift QSOs (see Smith *et al.* 1986), although this difference is not apparent in the near-infrared (Dunlop *et al.* 1993).

Clearly, our sample is too small to draw a firm conclusion and further observations will be required to establish the differences between radio-loud and radio-quiet host galaxies at high redshift. In particular, colour information will be vital, not only to establish conclusively that the extensions are due to the host galaxies, but also to provide some clues as to the evolutionary status of such systems.

## ACKNOWLEDGEMENTS

We would like to thank Chuck Steidel for providing us with his masking routines, and Mike Irwin for checking some of the subtraction results with his package IMAGES. IA thanks Eduardo Telles for his continuous help with IRAF. We are grateful to Max Pettini and Mike Irwin for providing useful comments on an earlier draft of this paper. IA's work is supported by the EEC HCM fellowship ERBCHBICT941023.